\documentclass[10pt]{article}
\usepackage[english]{babel}
\usepackage{amsfonts}
\usepackage[dvips]{graphicx}

\begin{document}

\title{\bf Modeling and ecodesigning crossflow ventilation fans with Mathematica}

\date{2010, July}

\author{\bf Gianluca Argentini \\
\normalsize{Research \& Development Dept., Riello Burners - Italy}\\
\normalsize gianluca.argentini@rielloburners.com \\
\normalsize gianluca.argentini@gmail.com \\}

\maketitle

\noindent{\bf Abstract}\\
The efficiency of a simple model of crossflow fan is maximized when the geometry depends on a design parameter. The flow field is numerically computed using a Galerkin method for solving a Poisson partial differential equation.\\

\noindent{\bf Keywords}\\
incompressible flow, stream function, differential problem, Galerkin method, impeller, fan.\\

\section{Crossflow fans}

\begin{figure}[ht!]
	\begin{center}
	\includegraphics[width=7cm]{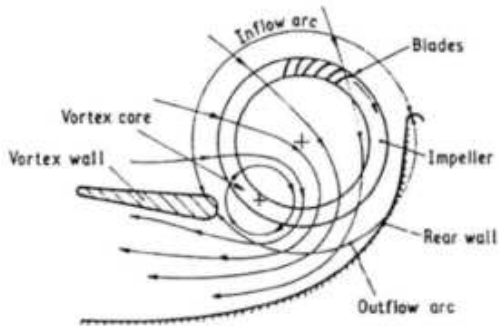}
	\caption{A crossflow fan schema.}
	\label{crossflow}
	\end{center}
\end{figure}

A crossflow, or tangential, fan (\cite{bleier}) is a ventilating structure where air is moved from an inlet zone to an outlet one using an impeller that forces the air to follow trajectories tangentially to the ideal circumferences concentrical to impeller itself, so that fluid goes across fan on planes that are normal to impeller rotation axes. The inlet and the outlet zone are physically separated by a fixed object, called {\it vortex wall} (see Fig.\ref{crossflow}).

Crossflow fans have the important property to get relatively high values of flow rate with small geometrical dimensions. Two typical engineering questions are how to design their housing and where to place the vortex wall between the inlet and the outlet zone (\cite{lazzaretto}). In this work, we try to get an answer to the second question using a simple two-dimensional geometry in the {\it Mathematica} numerical environment.

\section{The geometry}

\begin{figure}[ht!]
	\begin{center}
	\includegraphics[width=5cm]{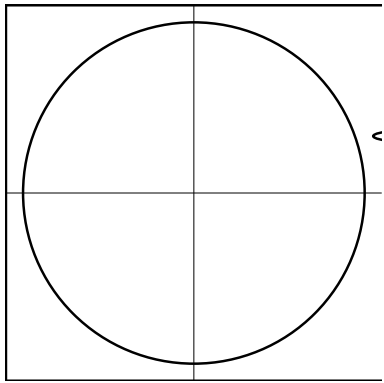}
	\caption{The geometry of fan.}
	\label{fan}
	\end{center}
\end{figure}

We consider a squared housing $D$ with edge of length $L$. The impeller is ideally represented by a circle of radius $R$, with $R < L$. In a cartesian coordinate system ${x,y}$, the left bottom vertex of the housing is the point $(-L/2,-L/2)$. The vortex wall is ideally represented by a point, vertex of a parabola, in the right edge of the square, at vertical position $s$, with $-L/2 < s < L/2$ (Fig.\ref{fan}). The segment $[s,L/2]$ is the inlet, the segment $[-L/2,s]$ is the outlet, so that the fluid particles have a counter-clokwise movement into the fan.

\section{Physics and Mathematics of fan}

Now we describe the mathematical model used for simulating the air-flow into the fan. We consider a zero-divergence velocity field $(u,v)$ in the 2D geometry, so that a scalar stream function $\Psi$ exists such that (see \cite{madani})

\begin{equation}\label{stream}
	u = \partial_y \Psi, \hspace{0.1cm} v = - \partial_x \Psi
\end{equation}

\begin{figure}[ht!]
	\begin{center}
	\includegraphics[width=3cm]{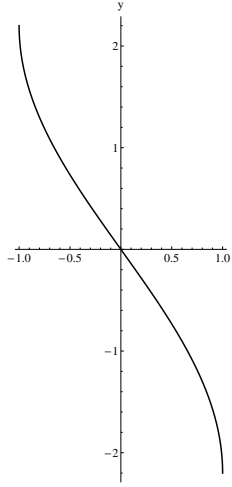}
	\caption{The $u_f$ profile for $L=4.4$ and $A=1$.}
	\label{ufComponent}
	\end{center}
\end{figure}

The expression $\partial_x v - \partial_y u = - \Delta \Psi$ is called {\it vorticity} (\cite{gurtin}). If the case of a simple counter-clokwise rotating field centered at the origin, we have $u = -w y, v = w x$ (\cite{madani}), so that $- \Delta \Psi = 2w$, where $w$ is the rotation angular speed. In the case of our crossflow fan, we can also considered for fluid particles a negative horizontal velocity component in the upper half $(0 \leq y \leq L/2)$ of the square and a positive one in the bottom half $(-L/2 \leq y \leq 0)$. In the case $s=0$, a simple way for describe this field $(u_f,v_f)$ is to consider the formulas

\begin{equation}\label{fanField}
	u_f = - A sin(\pi y/L), \hspace{0.1cm} v_f = 0
\end{equation}

\noindent where $A$ is a constant (see Fig.\ref{ufComponent}).

If $s \neq 0$, we use the correction $u_f = - A sin[\pi (s+y)/L]$. Therefore, in the case of an {\it estimated} field $(-w y + u_f, w x)$, the vorticity is $- \Delta \Psi = 2w -\pi A/L cos[\pi (s+y)/L]$. Note that, if $s = 0$ and $y$ small, using the approximation $cos\alpha \approx 1$ we have $2w = 2w \hspace{0.1cm} cos[\pi (s+y)/L]$, therefore we choose $A = 1$ (we are interested on comparison between geometries, not on absolute value of flow rate) and obtain the following simple expression for the vorticity of our crossflow fan:

\begin{equation}\label{vorticity}
	- \Delta \Psi = (2w - \pi/L)cos[\pi (s+y)/L]
\end{equation}

Previous formula is a differential equation of Poisson type (\cite{sneddon}) for the unknown function $\Psi$. On the three edges of the square that are walls for the fan we can impose the usual boundary conditions

\begin{equation}\label{boundaryConditions}
	\Psi(-L/2,y) = \Psi(x,-L/2) = \Psi(x,L/2) = 0
\end{equation}

\noindent because the normal component of flow along these edges is null (see \cite{madani}).

\section{Numerical resolution of Poisson equation}

If $s=0$, it is possible to try a symbolic resolution of previous differential problem, being the equation linear and nothing that a particular solution has the form $\Psi_0 = B cos[\pi (s+y)/L]$, with $B$ constant. For our purposes, we apply a version of the numerical method of Galerkin (\cite{schafer}), in the following fashion.
\noindent We seek for a solution in the form

\begin{equation}
	\Psi(x,y) = \sum_{i,j=1}^N c_{ij} \phi_{ij}(x,y)
\end{equation}

\noindent where $c_{ij}$ are constants to be computed and $\phi_{ij}$ are the linearly independent functions

\begin{equation}\label{testFunctions}
	\phi_{ij}=(x+L/2)(y+L/2)^i(y-L/2)^j
\end{equation}

\noindent which satisfy the boundary conditions (\ref{boundaryConditions}). The choice of these particular test-functions is justified by the fact that the flow is mainly determined by the position of the upper and bottom wall. Let $f(y)=2w -\pi A/L cos[\pi (s+y)/L]$. Then the Poisson equation can be approximated as

\begin{equation}
	- \sum_{i,j=1}^N c_{ij} \Delta \phi_{ij}= f
\end{equation}

\noindent Multiply the two members by $\phi_{nm}$ and integrate over the square $D$:

\begin{equation}\label{linearSistem}
	- \sum_{i,j=1}^N c_{ij} \int_D \left(\phi_{nm}\Delta \phi_{ij}\right) = \int_D \left(f \phi_{nm}\right)
\end{equation}

\noindent This is a system of $N^2$ linear algebraic equations in the unknowns $c_{ij}$ which can be solved by usual numerical methods.

\begin{figure}[ht!]
	\begin{center}
	\includegraphics[width=7cm]{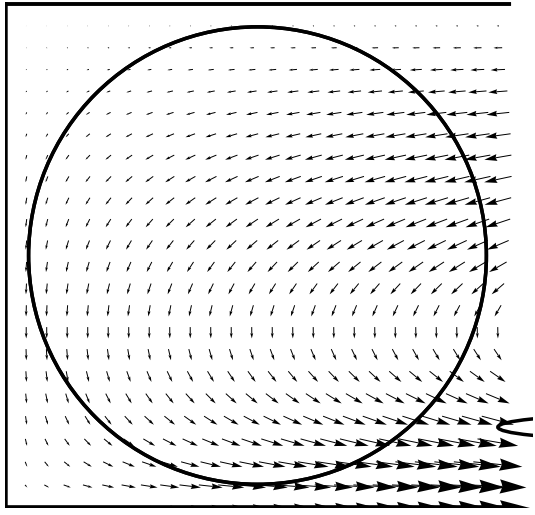}
	\caption{The velocity field for $pvw$=1.5 cm.}
	\label{field}
	\end{center}
\end{figure}

We have computed the coefficients $c_{ij}$ in the case $M=4$ using the methods built-in into the software {\it Mathematica}, ver. 7.0, for numerical integration and numerical resolution of linear systems. The geometrical and physical values which have been used for computations are impeller radius $R$ = 2.0 cm and rotation $rpm$ = 2000 ($w=2\pi rpm/60$), $L$ = 2($R$ + 0.2) cm, position of vortex wall $pvw$ such that $-1.5 \leq pvw \leq +1.5$ cm. A graphic result of the velocity field in the case $pvw$ = +1.5 is shown in Fig.\ref{field}.

\section{Ecodesign optimization}

In the last years European Union has emanated some Directives about energy related machines, with the purpose of regulating their design and utilization for ecological optimizations (\cite{ecodesign}). In particular, some studies has revealed the importance of optimize the design of ventilation fans for having high energetic efficiency (\cite{radgen}).

Suppose that the rotation speed of the impeller is constant for all the possible values of flow rate. This is only a first quite good approximation.
Then, the efficiency is determined by the product of the {\it flow rate} $Q$ and the {\it pressure head} $\Delta p$ of the fan. The pressure head is usually considered as the difference of the flow pressure at outlet and the flow pressure at inlet. The flow rate, defined as the dot product of the velocity field and the normal to the outlet surface, can be computed in our case with the formula

\begin{equation}\label{flowRate}
	Q = \int_{-L/2}^{pvw} u(L/2,y) dy
\end{equation}

\noindent Pressure $p$ and velocity field $\mathbf{v}$ are related by Navier-Stokes equations $\rho(\nabla \mathbf{v}) \mathbf{v} = - \nabla p$, where $\rho$ = 0.001 g/cm$^3$ is the value of air density, which we suppose to be constant along the flow. Once we have numerically computed $\mathbf{v}$, we can estimate numerically the pressure along the right vertical side of the fan casing by a path line integral $\int (\nabla p) \cdot d\mathbf{l}$ (see \cite{stewart}), so that if $a$ is the ordinate of a point e.g. in the outlet segment $[-L/2,pvw]$, we have

\begin{equation}
	p(a) = \int_a^{pvw} \partial_y p(L/2,y) dy
\end{equation}

\noindent For our purpose, we define the pressure head of the crossflow fan has the difference of the pressure at the middle point at outlet and the pressure at the middle point at inlet:

\begin{equation}\label{pressureHead}
	\Delta p = p[1/2(pvw-L/2)] - p[1/2(pvw+L/2)]
\end{equation}

\begin{figure}[ht!]
	\begin{center}
	\includegraphics[width=7cm]{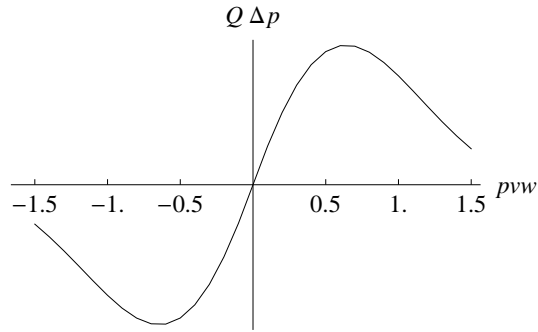}
	\caption{Graph of the product $Q \Delta p$.}
	\label{efficiency}
	\end{center}
\end{figure}

\noindent We have computed the efficiency product $Q \Delta p$ for values of $pvw$ in the interval $[-1.5,1.5]$. Using a step of $0.1$, the profile of the efficiency is shown in Fig.\ref{efficiency}.

\begin{figure}[ht!]
	\begin{center}
	\includegraphics[width=7cm]{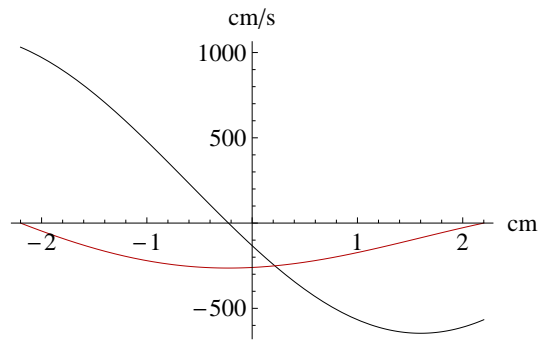}
	\caption{Profiles of $u$ (black) and $v$ (red) along the right side of the fan, in the case $pvw$ = 0.6 cm .}
	\label{velocityProfiles}
	\end{center}
\end{figure}

\noindent The maximum is obtained for $pvw$ = 0.6 cm, an information which were not easily guessed whitout a numerical analysis. From Fig.\ref{velocityProfiles}, we can note that, with this geometrical configuration, the speed at outlet is almost the double of the speed at inlet.

\end{document}